\documentclass[aps,prb,reprint,amssymb,amsmath,superscriptaddress,floatfix]{revtex4-1}

\usepackage{graphicx}
\usepackage{epstopdf}
\usepackage[english]{babel}
\bibpunct{[}{]}{,}{n}{}{}

\begin{document}

\title{Renormalization of the conduction band spectrum in HgTe quantum wells by electron-electron interaction}

\author{G.\,M.~Minkov}
\affiliation{School of Natural Sciences and Mathematics, Ural Federal University,
620002 Ekaterinburg, Russia}

\affiliation{M.N. Miheev Institute of Metal Physics of Ural Branch of Russian Academy of Sciences, 620137 Ekaterinburg, Russia}

\author{V.\,Ya.~Aleshkin}
\affiliation{Institute for Physics of Microstructures  RAS, 603950 Nizhny Novgorod, Russia}

\author{O.\,E.~Rut}
\affiliation{School of Natural Sciences and Mathematics, Ural Federal University,
620002 Ekaterinburg, Russia}

\author{A.\,A.~Sherstobitov}

\affiliation{School of Natural Sciences and Mathematics, Ural Federal University,
620002 Ekaterinburg, Russia}

\affiliation{M.N. Miheev Institute of Metal Physics of Ural Branch of Russian Academy of Sciences, 620137 Ekaterinburg, Russia}

\author{A.\,V.~Germanenko}

\affiliation{School of Natural Sciences and Mathematics, Ural Federal University,
620002 Ekaterinburg, Russia}

\author{S.\,A.~Dvoretski}

\affiliation{Institute of Semiconductor Physics RAS, 630090
Novosibirsk, Russia}

\author{N.\,N.~Mikhailov}

\affiliation{Institute of Semiconductor Physics RAS, 630090
Novosibirsk, Russia}
\affiliation{Novosibirsk State University, Novosibirsk 630090, Russia}

\date{\today}

\begin{abstract}
The energy spectrum of the conduction band in HgTe/Cd$_x$Hg$_{1-x}$Te quantum wells with a width $d=(4.6-20.2)$~nm has been experimentally studied in a wide range of electron density. For this purpose, the electron density dependence of the effective mass was measured by two methods: by analyzing the temperature dependence of the Shubnikov-de Haas oscillations and by means of the  quantum capacitance measurements. There was shown that the effective mass obtained for the structures  with $d<d_c$, where $d_c\simeq6.3$~nm is a critical width of quantum well corresponding to the Dirac-like energy spectrum, is close to the calculated values over the whole electron density range; with increasing width, at $d>(7-8)$~nm, the experimental effective mass becomes noticeably less than the calculated ones. This difference increases with the electron density decrease, i.e., with lowering the Fermi energy;  the maximal  difference between the theory and experiment  is achieved at $d = (15-18)$~nm, where the ratio between the calculated and  experimental masses reaches the  value of two and begins to decrease with a further $d$ increase. We  assume that observed behavior of the electron effective mass results from the spectrum renormalization due to electron-electron interaction.
\end{abstract}


\maketitle

\section{Introduction}
\label{sec:intr}

The quantum wells (QWs) in heterostructures HgTe/Cd$_x$Hg$_{1-x}$Te have a number of unusual properties compared to quantum wells based on  semiconductors with non-zero band gap. The reason for this is the ``negative'' band gap in the gapless semiconductor HgTe  in which $\Gamma_6$ band, forming a conduction band in conventional  semiconductors, is located below the $\Gamma_8$ band, which  forms the valence band. Such  arrangement of energy  bands leads to features of the energy spectrum of two-dimensional (2D) carriers, knowledge of which is required for reliable interpretation of all phenomena.

The energies of spatially quantized subbands at zero quasimomentum value, $k=0$,  and the energy spectrum $E(k)$ for different quantum well widths   ($d$) were calculated within $kP$ method in numerous papers (see, e.g., \cite{Gerchikov90,Zhang01,Novik05,Bernevig06,ZholudevPhD,Ren2016} and references therein). The different types of energy spectrum are realized with increasing width of the HgTe quantum well. Namely, it is normal when $d$ is less than a critical width $d_c\simeq 6.3$~nm, Dirac-like  at small quasimomentum for $d = d_c$ and inverted one when $d>d_c$, and finally the spectrum is semimetallic when $d>(14-16)$~nm. To interpret  experimental data, these calculations of the energy spectrum  are used practically always. They well describe the experimental data on the width dependence of the energies of  both  electron and hole  subbands at $k=0$ (see Ref.~\cite{Minkov17-1} and references therein).

However, quite a lot of differences between the experimental data and the results of the calculations has been accumulated  to date. First of all it refers to the spectrum of the top of valence band
\cite{Kozlov11,Kvon11-1,Minkov13,Olshanetsky12}. In the papers \cite{Minkov13,Minkov14,Minkov16}, we have shown that the interface inversion asymmetry (IIA) plays a crucial role in the forming of the energy spectrum of the valence band. Qualitative agreement between the experimental data and theory is achieved only when IIA is taken into account.

As for the energy spectrum of the conduction band, the theory predicts that the dispersion $E(k)$ is monotonic, nonparabolic and close to isotropic one for not very large $k$ values. This is surprising, but there was practically no experimental, consistent study of the dispersion law for the conduction band over a wide range of $d$ and energy.  Probably, this is the result of confidence in the $kP$ calculations and of a strong belief in that all the significant factors were taken into account in them.

\begin{figure}
\includegraphics[width=0.95\linewidth,clip=true]{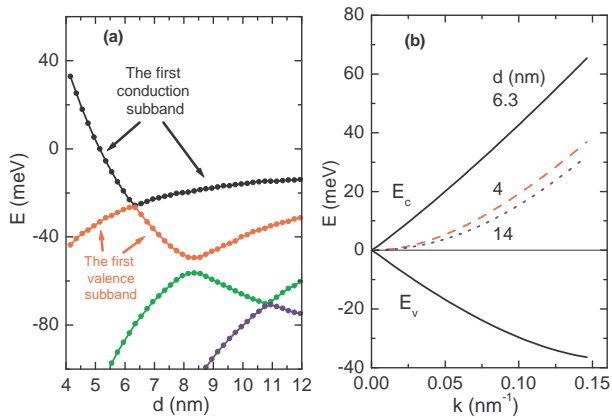}
\caption{(Color online) The quantum well width dependence of the subband energies at $k=0$ (a) and $E$~vs~$k$ dependences for some QW widths (b). The $E$~vs~$k$ dependence for the valence band presented for critical width shows that linear dependence is realized only for small quasimomentum.}
\label{F1}
\end{figure}

The study of the effective  mass of carriers is a powerful tool to probe the energy spectrum and its  evolution with variations of quantum well width and carrier density.

\begin{figure}
\includegraphics[width=1.0\linewidth,clip=true]{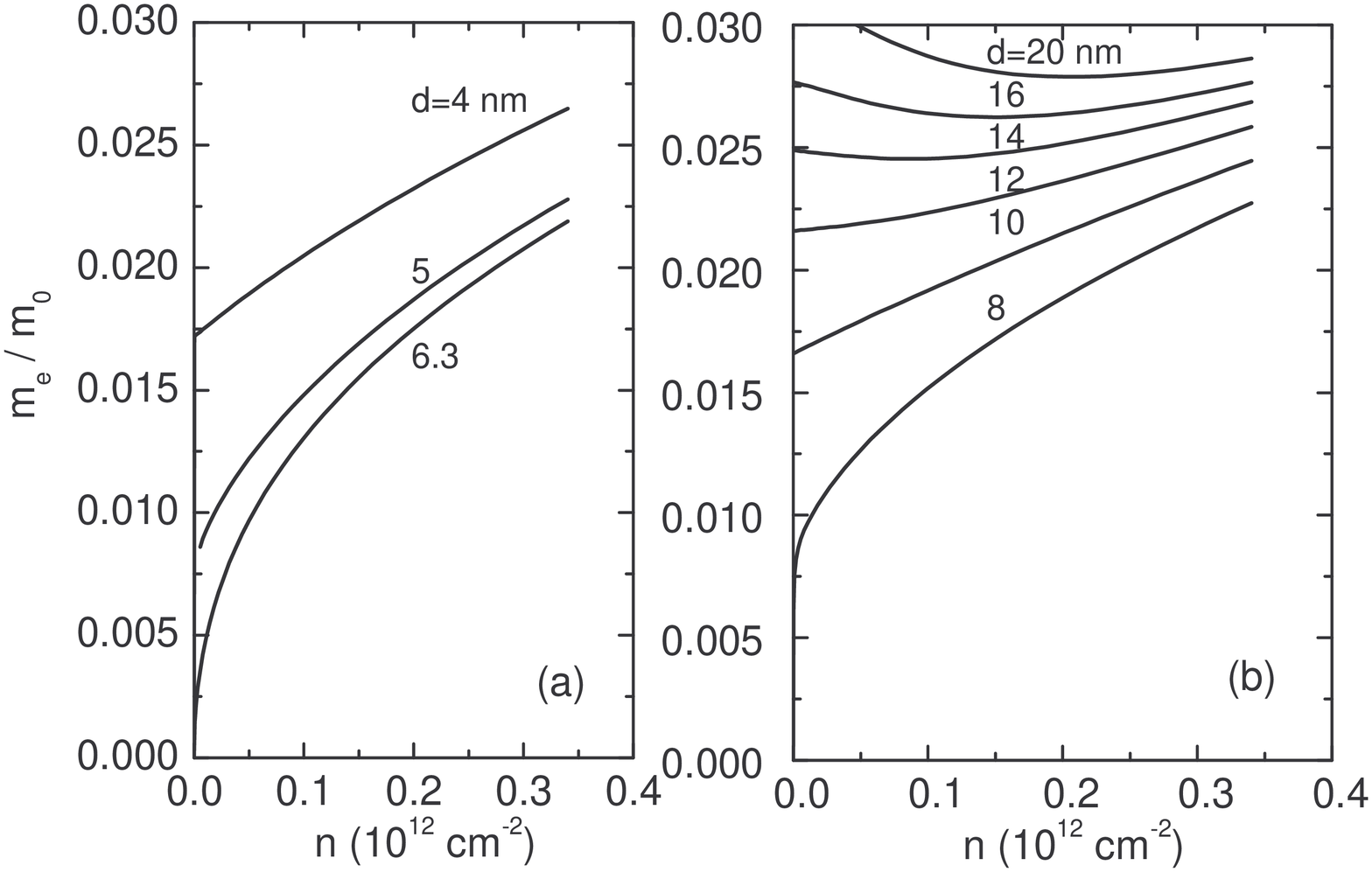}
\caption{(Color online) The calculated dependences of the electron effective mass on the electron density for  $d\leq d_c$ (a) and for $d>d_c$ (b).}
\label{F2}
\end{figure}

Let us consider what the theory predicts for the effective mass of electrons in the conduction band. We have calculated the energy spectrum for the HgTe quantum wells grown on the $(013)$ plane  within the framework of four-band \emph{kP} model \footnote{We used the parameters from Ref.~\cite{Novik05}.}. The results for some QW widths are shown in Fig.~\ref{F1}. The energies of the  2D subbands at $k=0$  are presented in Fig.~\ref{F1}(a), while the dispersion $E(k)$ is shown in Fig.~\ref{F1}(b). Because the energy spectrum of the conduction band strongly  nonparabolic,  the electron effective mass ($m$) depends not only on the quantum well width but on the energy also. Therefore in Fig.~\ref{F2} we plot the electron effective mass, $m=\hbar^2 k (dE/dk)^{-1}$, as a function of electron density ($n$) for quantum wells of  different width,  $d\leq d_c$ and $d > d_c$.
One can see that at $d<14$~nm the effective mass increases with grown density monotonically,  while at $d>14$~nm $m$ decreases at low $n\lesssim 1\times 10^{11}$~cm$^{-2}$ and starts to increase at higher  $n$ values.

To our knowledge, the detailed measurements of $m(n)$ were performed only by means of the cyclotron resonance for $d=7$~nm and $8$~nm \cite{Ikonnikov11}. The data obtained turn out lower by ($10-15$) percent as compared with the calculated results. The available data for wells of other widths are fragmentary, which makes it impossible to understand whether these results are consistent with the calculations or not.

In this paper we present the results of detailed study of electron effective mass within the wide electron density range $n=(1-5)\times 10^{11}$~cm$^{-2}$  in the structures with the different QW width, $d=(4.6-20.2)$~nm. The  electron effective mass is determined by analyzing the temperature dependences of the amplitude of the Shubnikov-de Haas (SdH) oscillations and from the quantum capacitance measurements. The data obtained are interpreted within the framework of the four-band $kP$ model. We show that  the experimental values of the effective mass  are close to the calculated ones over the whole density range in the structures  with $d<d_c$. With the increasing QW width the experimental values of effective mass become noticeably less than the calculated ones. The maximal  difference between the theory and experiment  is achieved at $d = (15-18)$~nm, where the ratio between them  reaches the value of two. Our assumption is that the observed behavior of the electron effective mass results from the spectrum renormalization due to many-body effects.

\section{Experiment}
\label{sec:expdet}

\begin{figure}
\includegraphics[width=0.8\linewidth,clip=true]{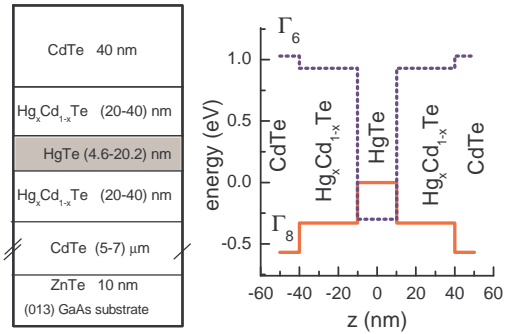}
\caption{(Color online) Sketch and energy diagram of the structures investigated. }
\label{F3}
\end{figure}

Our HgTe quantum wells were realized on the basis of HgTe/Cd$_{x}$Hg$_{1-x}$Te ($x=0.5-0.7$)  heterostructure grown by molecular beam epitaxy on GaAs substrate with the ($013$) and ($100$) surface orientation \cite{Mikhailov06}. The nominal widths of the quantum wells under study were $d = (4.6-20.2)$~nm. The samples were mesa etched into standard Hall bars of $0.5$~mm width with the distance between the potential probes of $0.5$~mm. To change and control the electron densities  in the quantum well, the field-effect transistors were fabricated with parylene as an insulator and aluminium as a gate electrode. The measurements were performed at the temperature  $T=(4 - 12)$~K in the  magnetic field up to $2.0$~T. For each heterostructure, several samples were fabricated and studied. The sketch and the energy diagram of the structures investigated are shown in  Fig.~\ref{F3}. The main parameters are listed in Table~\ref{tab1}.

\begin{table}
\caption{The parameters of  heterostructures under study}
\label{tab1}
\begin{ruledtabular}
\begin{tabular}{ccccc}
number & structure & $d$ (nm) & type & $\mu^\text{a}(10^5\text{~cm}^{2}/$Vs)\\
\colrule
  1& 150220 & 4.6    & $n$   & 0.39  \\
  2& 081110 &  5      & $n$    & 0.22   \\
  3& 160126 & 5      & $p$   & 0.39  \\
  4& 071225 & 8.3     & $p$   & 1.2 \\
  5& 081107 & 9.5      & $p$   &1.1  \\
  6& 150224 & 10   & $n$   &2.1   \\
  7& 130412 & 15    & $p$   &3.4   \\
  8& 100623 & 18    & $p$   &6.8   \\
  9& 110614 & 20.2  & $p$   &4.5   \\
\end{tabular}
\end{ruledtabular}
\footnotetext[1]{After illumination, at $n=2\times 10^{11}$~cm$^{-2}$. }.
\end{table}

\section{Results and discussion}
To study the energy spectrum  we have measured the dependence of the  effective mass of electrons on their density. The values of $m$ was determined by two methods: by measuring the  temperature dependence of the amplitude of the SdH oscillations  and by analysing the electron density dependence of the quantum capacitance. Such measurements were performed for all the structures. The experimental results and their analysis are similar ,  therefore, as an example, let us consider in more detail the data obtained for the structure 110614 with  $d=20.2 $~nm.

\begin{figure}
\includegraphics[width=0.95\linewidth,clip=true]{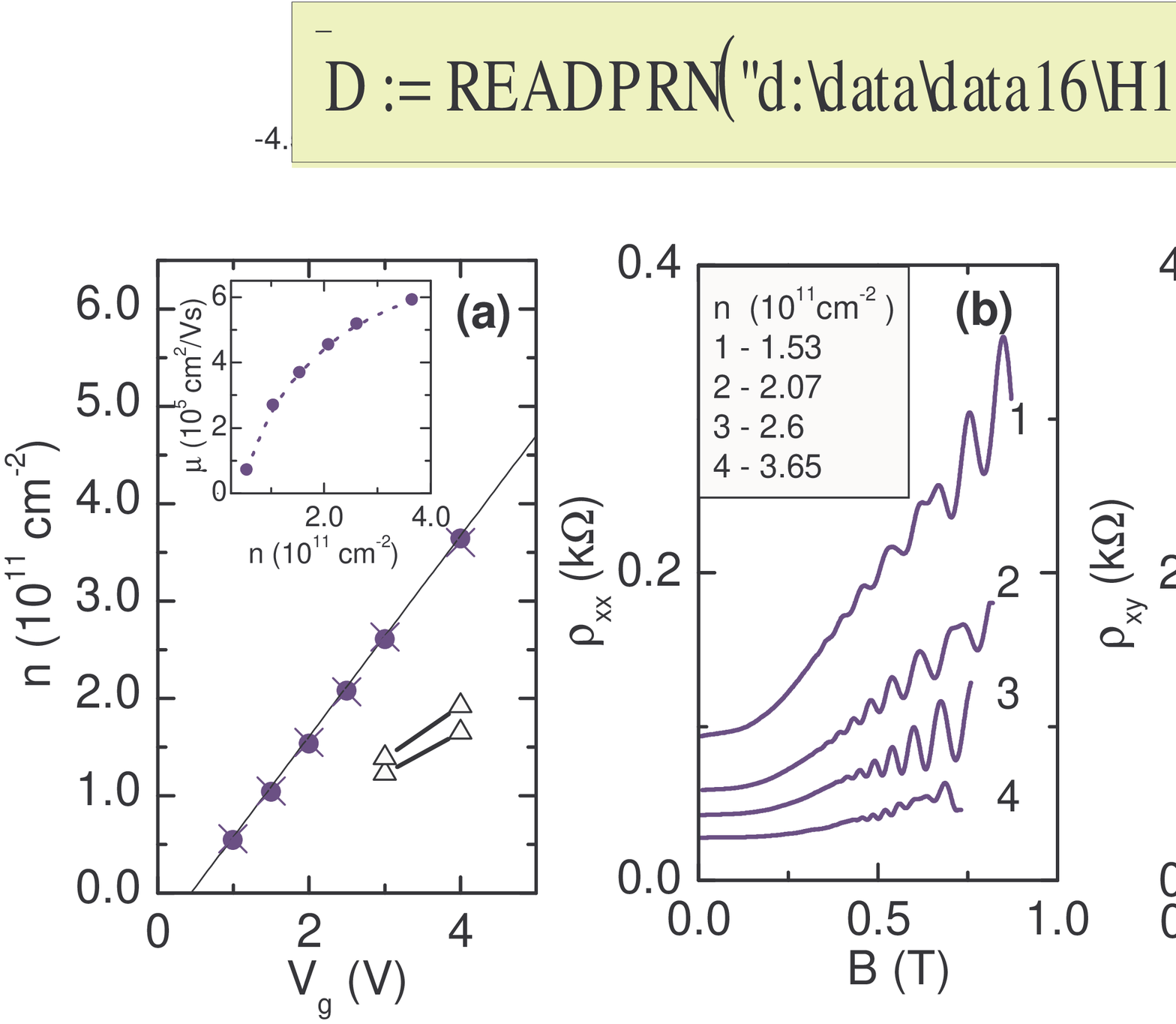}
\caption{(Color online) (a) -- The gate voltage dependence of the electron density found from the Hall effect as $n_H=-1/eR_H(0.2\text{~T})$ (circles) and from the frequency of the SdH oscillations (crosses)  (see the text,  for more details). The triangles show the electron densities in the each of SO split subbands. The inset  shows the density dependence of the Hall mobility. (b) and (c) -- The magnetic field dependences of  $\rho_{xx}$ and $\rho_{xy}$, respectively,  for some electron densities. Structure 110624. $T=4.2$~K.}
\label{F4}
\end{figure}

To determine the basic parameters of the structures, we have measured the dependences of the longitudinal ($\rho_{xx}$) and transverse ($\rho_{xy}$) magnetoresistance on a magnetic field ($B$) for different gate voltages ($V_g$). The results are shown in  Fig.~\ref{F4}.

These measurements show that the electron density obtained from the Hall effect as $n_H=-1/eR_H(0.2\text{~T})$  depends on the gate voltage linearly,
$n_H(V_g)= -4.5\times 10^{10}+dn/dV_g\times V_g$,~cm$^{-2}$ with $dn/dV_g=1.03\times 10^{11}$ cm$^{-2}$V$^{-1}$ [see Fig.~\ref{F4}(a)].  This value of $dn/dV_g$ is in a good agreement with $dn/dV_g$ found from the capacitance measurement $dn/dV_g=C/eS_g$, where $C$ is the capacitance between the 2D gas and the gate electrode, measured for the same structure, $S_g$ is the gated area, $e$ is the elementary charge. The inset in Fig.~\ref{F4}(a) shows that the electron mobility increases with the growing density and achieves the value of $\mu\simeq 6\times 10^5$ cm$^2$/Vs at $n\simeq 4\times 10^{11}$~cm$^{-2}$.

One can see from Fig.~\ref{F4}(b) that  the SdH oscillations  are  observed on the background of  positive magnetoresistance \footnote{The monotone magnetoresistance in 2D systems was discussed in numerous paper,  see, e.g., Refs.~\cite{Dmitriev08-1,Dmitriev12} and references therein. This issue is beyond the scope of the present paper.}. An inspection of  the data  shows that two different regimes in the SdH-oscillations picture can be clearly recognized.
The first one corresponds to a relatively low electron density, $n<2.5\times 10^{11}$~cm$^{-2}$, where the oscillations are unsplit at low magnetic fields, $B<(0.4-0.6)$~T. The electron density found from the period of SdH oscillations supposing two-fold degeneracy of the Landau levels (LLs) coincides with $n_H$ [see Fig.\ref{F4}(a)]. In the higher magnetic fields, $B>(0.4-0.6)$~T, the Zeeman splitting of the oscillations starts to be observed.

In second regime, of the higher  electron densities, $n>2.5\times 10^{11}$~cm$^{-2}$, the SdH oscillations differ drastically. The beating of oscillations resulting from the Bychkov-Rashba spin-orbit (SO) splitting  become clearly evident. Therewith, the Fourier spectra exhibit two maxima corresponding to these split subbands that allows us to obtain the electron density in each of them [shown by the triangles in Fig.~\ref{F4}(a)].  The sum of these densities shown  in Fig.~\ref{F4}(a) by the crosses  coincides  well with $n_H$ also.

\begin{figure}
\includegraphics[width=1\linewidth,clip=true]{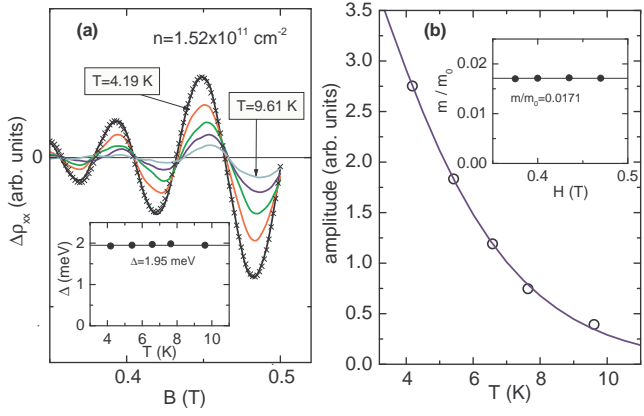}
\caption{(Color online) (a) -- The oscillating part  of $\rho_{xx}$ taken at different temperatures for $n=1.47\times 10^{11}$ cm$^{-2}$. The crosses show the results of the best fit by Eq.~(\ref{eq01}) of the curve at $T=4.19$~K. The inset shows the broadening parameter $\Delta$ found from the magnetic field dependences of the oscillation amplitude at different $T$. (b) -- The experimental  temperature dependence of the amplitude of the SdH oscillations at $B=0.45$~T (circles) and the best fit by Eq.~(\ref{eq02}) (curve). The inset presents the values of $m$ found at different $B$. }
\label{F5}
\end{figure}

To determine the electron effective mass let us first analyze the  data for the low electron density when neither the Zeeman nor SO splitting of the oscillations is observed. The oscillating part of $\rho_{xx}(B)$ for the different temperatures  obtained by extracting the monotonic part for the particular case of $n = 1.47\times 10$~cm$^{-2}$ is shown in Fig.~\ref{F5}(a). These oscillations are well described by the  Lifshitz-Kosevich formula which for the unsplit oscillations reads
\begin{equation}
\frac{\rho_{xx}(B,T)-\rho_{xx}(0,T)}{\rho_{xx}(0,T)}\propto A(B,T)\,\cos{\left(\frac{2\pi F_{1/B}}{B}\right)}
\label{eq01}
\end{equation}
with
\begin{equation}
A(B,T)=\exp\left( -\frac{\Delta}{\hbar \omega_c}\right)\, \frac{2\pi ^2 k_BT}{\hbar \omega_c}{\Big \slash}\sinh\left(\frac{2\pi ^2 k_BT}{\hbar \omega_c}\right),
\label{eq02}
\end{equation}
where $F_{1/B}$ is the oscillation frequency in the reciprocal magnetic field,  $\omega_c=eB/m$,  $k_B$ is the Boltzmann constant, $\Delta$ is the broadening of the Landau levels.  This is demonstrated by  Fig.~\ref{F5}(a), where an example of the best fit by Eq.~(\ref{eq01}) for $T=4.19$~K is presented by the crosses. As seen the fitting result  coincides with the experimental curve very closely.

The temperature dependence of the oscillation amplitude at $B=0.45$~T [see Fig.~\ref{F5}(b)]  is well described by Eq.~(\ref{eq02}) that gives the electron effective mass $m=(0.0171\pm 0.0005) m_0$. The inset in Fig.~\ref{F5}(b) shows that the values of the mass found at different magnetic fields are very close to each other.

Such a method of finding the effective mass assumes that the broadening parameter $\Delta$ does not depend on the temperature. To verify that this condition is fulfilled, the values of $\Delta$ were determined from the magnetic field dependences of the amplitude of the oscillations measured at different temperatures. The results of such analysis shown in the inset in Fig.~\ref{F5}(a) demonstrate that $\Delta$ is really independent of $T$ in our case \footnote{Recent trend is the use of the global fit of the oscillation curves measured at different temperatures by the theoretical expression like Eq.~(\ref{eq01}) to obtain the spectrum parameters. The procedure used in the present paper is more controllable and evident in our opinion.}.

To find the effective mass over a wide range of the electron density it is more suitable to measure the oscillations of $\rho_{xx}$ at a fixed magnetic field as a function of gate voltage rather than swiping the magnetic field at a fixed gate voltage.   Because the  resistivity strongly varies with changing $V_g$ (especially at low electron density), the measurements of the first or even second derivative of $\rho_{xx}$ with respect to $V_g$ are more convenient  to study the  oscillations of low amplitude. To do this an \emph{ac} gate voltage $\delta V_g^{ac}$=20~mV (that corresponds to $\delta n\simeq 2\times10^9$~cm$^{-2}$)  with  the frequency $f=37$~Hz was applied together with \emph{dc} voltage  $V_g$. It is obviously, that such a  modulation of gate voltage results in the modulation of voltage drop  $\delta V_{xx}^{ac}$ between the potential probes. So, registering the signal between the potential probes  on the  frequencies $f$ and $2f$ we measure actually the value which is proportional to $d\rho_{xx}/dV_g$ and $d^2\rho_{xx}/dV_g^2$, respectively. It is easy to show that the temperature dependences of  $d\rho_{xx}/dV_g$ and $d^2\rho_{xx}/dV_g^2$ are described by the same equation, Eq.~(\ref{eq02}), as the oscillations of $\rho_{xx}$ and, hence, the effective mass can be determined by the same  manner.

\begin{figure}
\includegraphics[width=1\linewidth,clip=true]{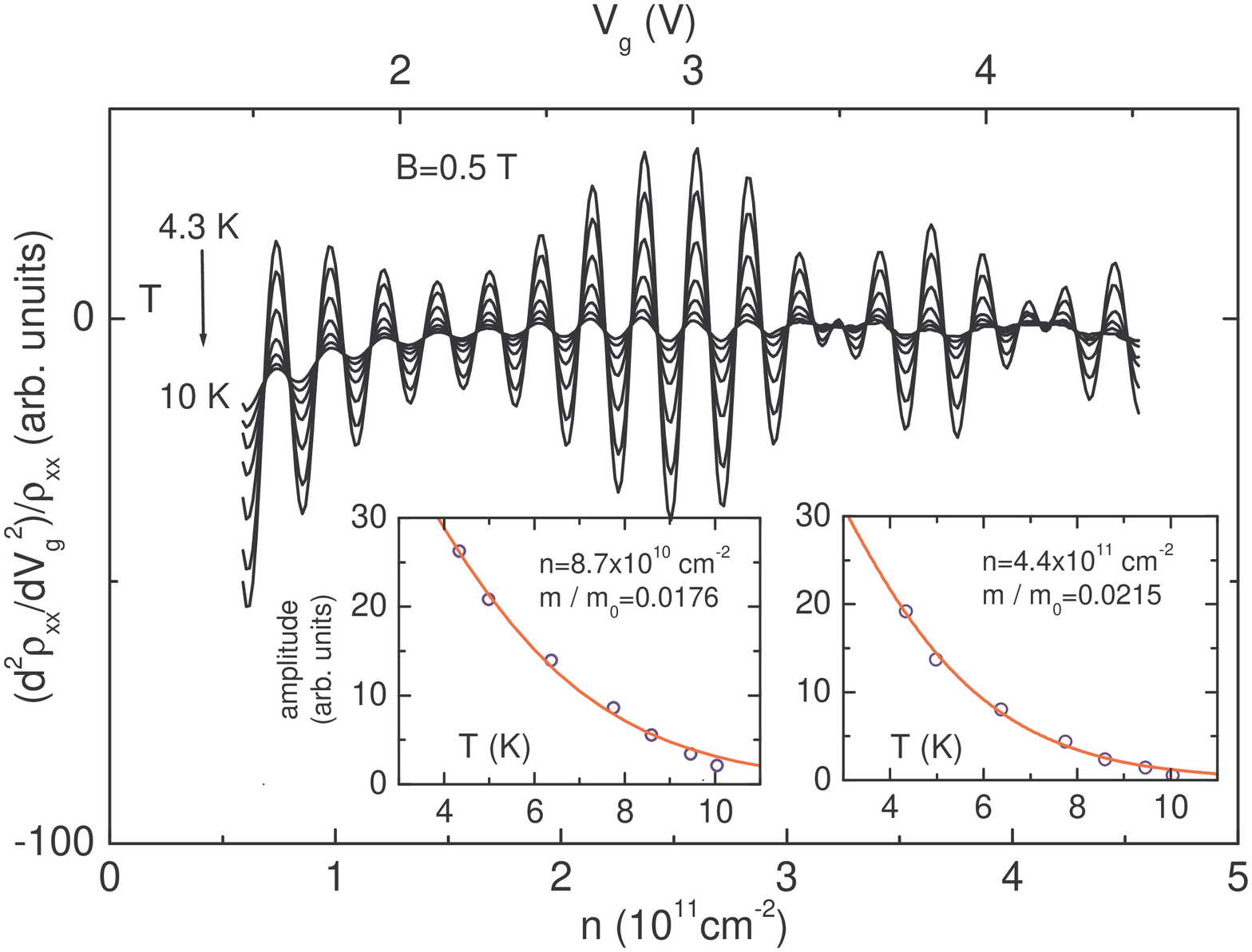}
\caption{(Color online) The gate voltage dependences of  $d^2\rho_{xx}/dV_g^2$ at $B=0.5$~T for different temperatures. The insets show the temperature dependences of the oscillations amplitudes for two electron densities. The symbols are the data, the curves are the results of the best fit by Eq.~(\ref{eq02}) with $m=0.0176\,m_0$ and $0.0215\,m_0$.}
\label{F6}
\end{figure}

The gate voltage dependences of  $d^2\rho_{xx}/dV_g^2$ measured  at  $B=0.5$~T and  different temperatures are shown in Fig.~\ref{F6}. As seen these oscillations are near-periodic in $V_g$. This is easy to understand because to change the oscillation number by one at fixed $B$  it is needed to change the electron density by the value equal to the LL degeneracy which with taking into account the ``spin'' degeneracy is $2Be/h$.

Some nonperiodicity arises from the SO splitting, which increases with increasing $n$ and leads to beating of oscillations, obvious at $V_g>3$~V.  The presence of the beating does not interfere with the possibility to determine the effective mass from the temperature dependence of the oscillation amplitude in low magnetic fields where the oscillations are pairwise merged, i.e., at $V_g$ far from the nodes. Indeed, it is easy to show that  in this case the energy distance between pairwise merged oscillations is equal to $\hbar\omega _c=\hbar eB/m^{av}$ with $m^{av}=(m^{-}+m^{+})/2$,
where $m^{-}$ and $m^{+}$ are the electron effective masses for SO split subbands. This relation remains valid, regardless of which levels have merged; $N^+$ and $N^-$ or $N^+$ and $(N^-+1)$. The insets in Fig.~\ref{F6} show the experimental temperature dependences of the oscillation amplitude for two electron densities and the results of the best fit of these data by Eq.~(\ref{eq02}). As seen Eq.~(\ref{eq02}) describes the experimental data nicely.

\begin{figure}
\includegraphics[width=0.75\linewidth,clip=true]{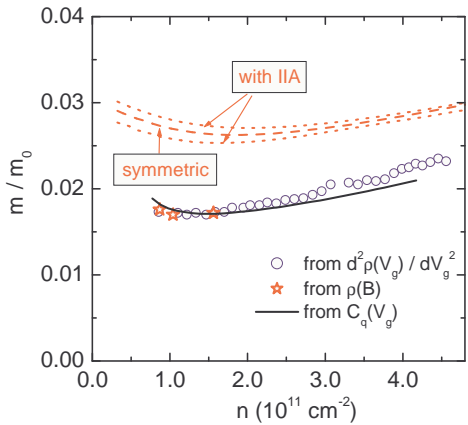}
\caption{(Color online) The electron effective mass found from the temperature dependences of the amplitude of the SdH oscillations of $\rho_{xx}$ at fixed gate voltage (stars), from the temperature dependences of the oscillation amplitude of $d^2\rho_{xx}/dV_g^2$ at fixed $B$ (circles) and from the gate voltage dependence of $C_q$ at $B=0$ (solid line).
The dashed and dotted lines are calculated withou and with taking into account interface inversion asymmetry.}
\label{F7}
\end{figure}

All the data on the effective mass of electrons obtained by these two methods in the entire density range are collected in Fig.~\ref{F7}. One can see that the $m$ values  are in a good agreement with each other and effective mass demonstrates a $1.5$-times increase over the density range
$(0.7-4.5)\times 10^{11}$~cm$^{-2}$.

Before  comparing these data with the theory let us consider the experimental results obtained by another method,  namely found from the gate voltage dependence of the quantum capacitance ($C_q$) measured for the same structure. The specific capacitance between the gate electrode and 2D gas is
\begin{equation}
1/C=1/C_g+1/C_q,\,\,\,     C_q=e^2D,
\label{eq03}
\end{equation}
where $C_g$ is geometrical capacitance, $D$ is the density of states of 2D gas which is related to the effective mass of the carriers as $D=m/(\pi\hbar^2 )$.

\begin{figure}
\includegraphics[width=0.87\linewidth,clip=true]{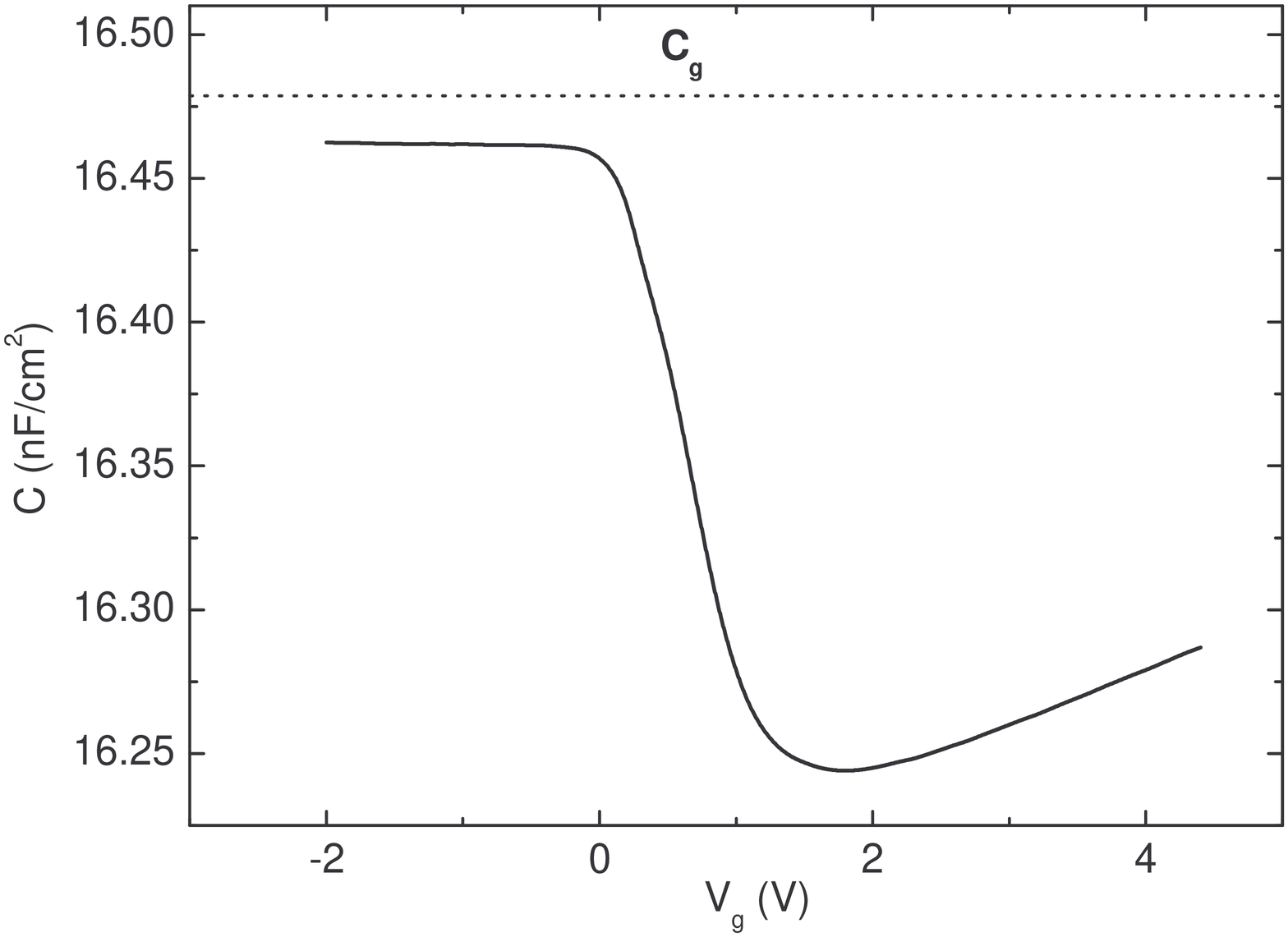}
\caption{(Color online) The gate voltage dependence of the capacitance between the gate electrode and 2D gas. The dashed line shows the geometric capacitance. $T=4.2$~K. }
\label{F8}
\end{figure}

The experimental gate voltage dependence of the capacitance is presented in Fig.~\ref{F8}. To find the electron effective mass from this data in accordance with Eq.~(\ref{eq03}) as
\begin{equation}
\frac{m}{m_0}= \frac{\pi\hbar^2}{e^2m_0}\frac{C}{1-C/C_g}
\label{eq04}
\end{equation}
one should know the geometrical capacitance. The accuracy of direct determination of the specific capacitance is not sufficient. It is limited by the accuracy of the gate area measurement. The value of $C_g$ was determined in such a way that the density of states of the top of the valence band corresponded to the two-fold degenerate states with the effective mass  of $(0.25\pm 0.02) m_0$. Such a mass value was experimentally found in Ref.~\cite{Minkov17} in structures with a quantum well width of $(8-20)$~nm. The result with $C_g=16.48$~nF/cm$^2$ is plotted in Fig.~\ref{F7}. Note, the variation of the hole effective mass within this range gives very small variation in determination of electron effective mass.

Fig.~\ref{F7} shows that all the data on the effective mass  obtained  by the three  different methods are well matched. Some growth of the effective mass found from $C(V_g)$ with the density decrease  at $n<8\times 10^{10}$~cm$^{-2}$  results most probably from an additional density of states of a tail of the valence band at these energies.

Now we are in position to compare the experimental data with theoretical results. The electron density dependences of the electron effective mass calculated within a four-band $kP$ method for symmetrical (013)-HgTe QW with  $d=20.2$~nm are plotted in Fig.~\ref{F7} by the dashed  line. It is seen that the theoretical curve lies significantly above the experimental data.
At a low electron density, the theoretical effective mass is approximately $50$~percent greater than the experimental one. This difference diminishes with increasing electron density and at $n=4.5\times 10^{11}$~cm$^{-2}$ it does not exceed $25$~percent.

It is known that interface inversion asymmetry leads to a strong SO splitting and significant reconstruction of the valence band spectrum \cite{Minkov13,Minkov14,Minkov16}.
Therefore, in Fig.~\ref{F7}, we present the $m$~versus~$n$ dependences calculated with taking into account IIA with parameters from Ref.~\cite{Minkov17}.  It can be seen that IIA leads to a small correction to the electron effective mass over the whole experimentally accessible region of the electron density and does not improve accord with the experimental data.

Thus, the electron effective mass obtained experimentally at low densities is significantly smaller than that calculated within the framework of the \emph{kP} model.

Such measurements and data treatments have been performed for all the structures listed in the Table~\ref{tab1} and  now we briefly consider the results for the structure 150220  with narrow quantum well, $d = 4.6$~nm ($d <d_c$).

\begin{figure}
\includegraphics[width=0.75\linewidth,clip=true]{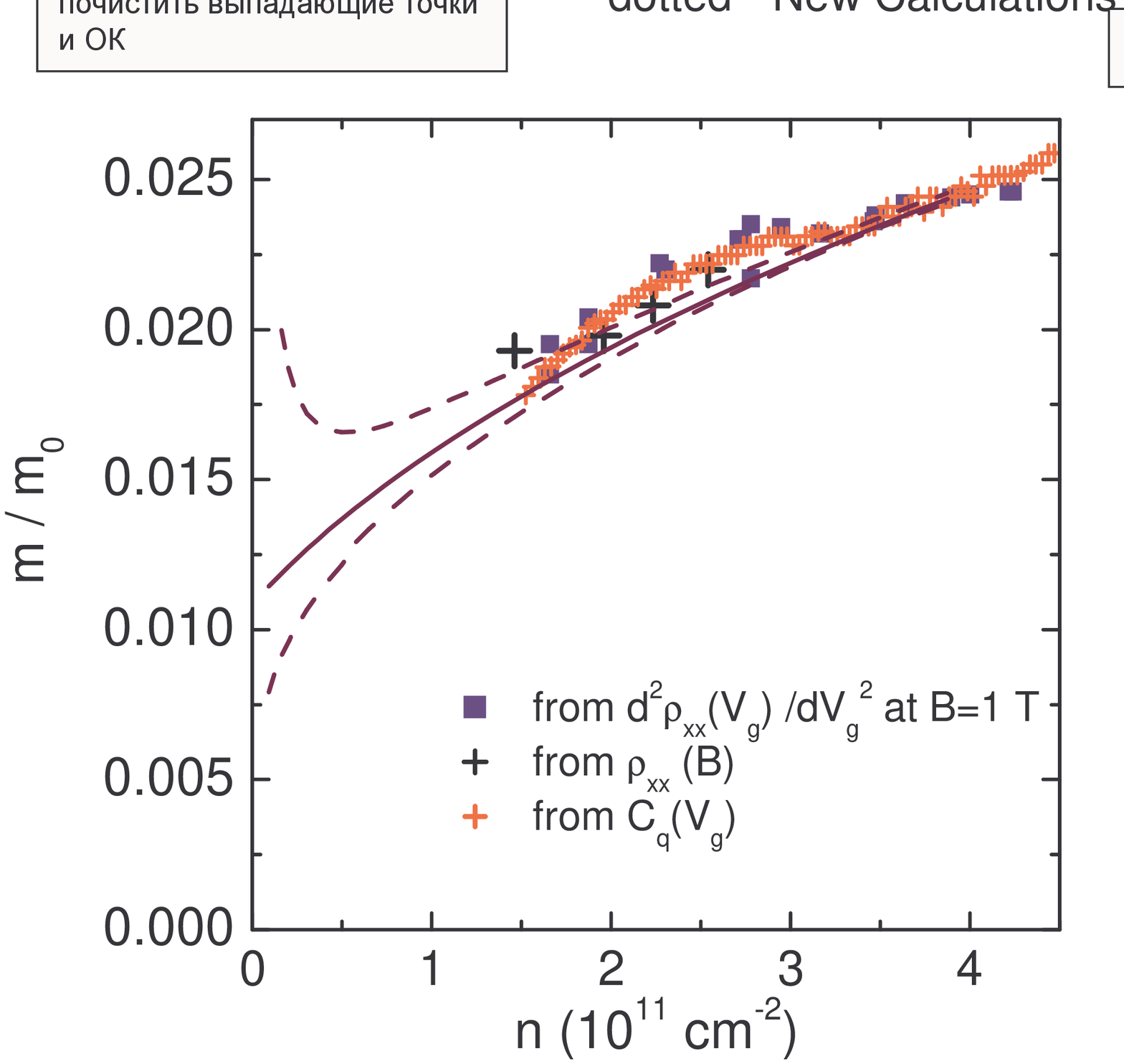}
\caption{(Color online) The electron density dependence of the electron effective mass. The symbols are the experimental data obtained by three methods as described in the text. The solid and dashed curves are the theoretical results calculated without and with taking into account IIA, respectively. Two dashed lines correspond to two branches split by SO interaction. Structure 150220.
}
\label{F9}
\end{figure}

The electron density dependence of the effective mass  obtained by all three methods are presented in Fig.~\ref{F9}. As seen all the data are consistent with each other as well. The theoretical  $m$~vs~$n$ curves calculated without and with taking into account IIA for $d = 4.6$~nm are also presented in Fig.~\ref{F9}. Unlike the data for $d=20.2$~nm (Fig.~\ref{F7}), the experimental  results are very close to theoretical ones.

\begin{figure}
\includegraphics[width=1\linewidth,clip=true]{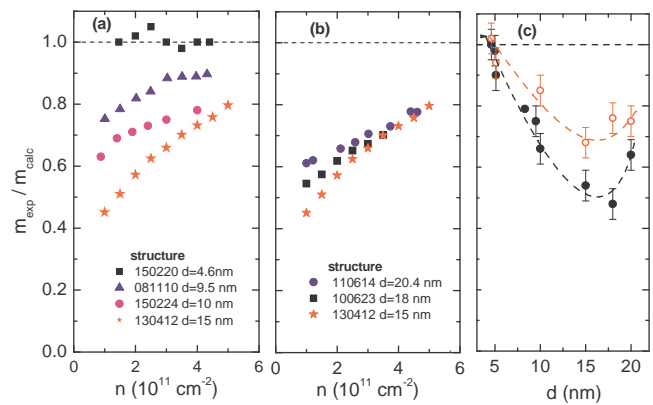}
\caption{(Color online) The dependences of the ratio $m_{\text{exp}}/m_\text{calc}$ as a function of the density for $d\leq 15$~nm (a) and for  $d\geq 15$~nm (b). (c) -- The values of $m_{\text{exp}}/m_\text{calc}$ plotted as a function of QW width for electron densities $2\times 10^{11}$~cm$^{-2}$ (solid circles) and $4\times 10^{11}$~cm$^{-2}$ (open circles).}
\label{F10}
\end{figure}

In order to consider how the theory describes the experimental data in all the structures with different QW width from $4.6$~nm to $20.2$~nm over the  whole electron density range,  we have plotted  the ratio of experimental effective mass to the theoretical one ($m_{\text{exp}}/m_\text{calc}$) as a function of the electron density  for most of
structures investigated in  Figs.~\ref{F10}(a) and \ref{F10}(b). In Fig.~\ref{F10}(c), this ratio is depicted as a function of the QW width for two electron densities. Inspection of Fig.~\ref{F10} shows that: (i) in the structures  with $d<d_c$, the experimental values of effective mass  are close to the calculated ones over the whole density range; (ii) with increasing QW width, at $d>(7-8)$~nm, the experimental  effective mass becomes noticeably less than the calculated ones. This difference increases with the electron density decrease, i.e., with the  lowering Fermi energy; (iii) the maximal  difference between the theory and experiment  is achieved at $d = (15-18)$~nm, where the $m_\text{calc}$ to $m_\text{exp}$ ratio reaches the  value of two and begins to decrease with a further $d$ increase.

What are possible reasons for this discrepancy between the experimental data and the generally accepted theory for the electron energy spectrum in HgTe quantum wells?

The first group of reasons can be associated with inaccuracies in the parameters used in the calculation. Among these parameters are the width of the quantum well, the value of deformation due to a lattice mismatch between the QW and barrier materials, the possible presence of cadmium in the quantum  well, the bands offsets at the QW interfaces, the different values of the matrix element of the momentum operator in the barriers and quantum well. The variation of these parameters in the calculations shows that the  small value of the electron mass observed experimentally at $d=(13-18)$~nm can be obtained only by assuming presence of $5$ percent of cadmium in the well or a 20 percent increase in the value of matrix element of momentum operator $P$.

However, these assumptions lead to  strong discrepancies with the other experimentally measured effects. In particular, investigating the QW width dependence of the resistivity in the charge neutrality point we have observed the sharp deep minimum at $d=(6.0\pm 0.5)$~nm corresponding the the critical width $d_c$ in which the Dirac-like gapless spectrum is realized. With the $5$ percent cadmium in the quantum well or $20$ percent increase in the value of the matrix element $P$, the minimum should be observed at $d\simeq 10$~nm.

Another possibility to check that the cadmium is absent in the quantum well is based on the specific feature of the energy spectrum quantization in the magnetic field. It is well known that there are two anomaly Landau levels in the structures with inverted spectrum \cite{Schultz98,Minkov13,Koenig07}. They start at $B = 0$ from the bottom of the conduction band and the top of the valence band, and moving toward each other with the growing magnetic field they cross each other in the magnetic field $B = B_c$. The $B_c$ value strongly depends on the parameters mentioned above. Experimentally, the cross is observed namely at those $B_c$ which are predicted by the $kP$ calculations for HgTe quantum well with the use of the conventional set of  parameters.

In addition to the above-mentioned ``technological'' factors, there are physical reasons. First of all, this may be the role of many-body effects, which were not taken into account in the framework of the used \emph{kP} calculations. It is well known that the exchange part of the Coulomb interaction leads to an increase in the effective mass in the single-band approximation, i.e., to the opposite change in $m$ value as compared with that observed in our case. Role of the exchange interaction can differ drastically for narrow-gap or gapless 2D systems.
For such systems, one should take into  account both inter- and intra-band contributions which have different signs \cite{Kusminskiy}.

As for experiments, the interaction effects reveal themselves most conspicuous in graphene. So in Ref.~\cite{Elias2011}, a strong reshape of the Dirac cones due to the interaction effect in suspended graphene has been investigated by means of SdH oscillations. It was shown that many-body effects lead to a decrease in the effective mass by the factor of  $(2.0-2.5)$ at low densities, and this decrease agrees well with the calculations made for a simple, initially linear spectrum.

We suppose that namely many-body  effects lead to the decrease of effective mass shown in Fig.~\ref{F10}. Note in this paper we present the  results for the structures with the highest mobility which we have.  The decrease in the effective mass in structures with lower mobility is noticeably less that is in a qualitative agreement with the theoretical prediction according to which a disorder  kills the exchange contribution to energy spectrum \cite{Lozovik}. Now we cannot systematically study the role of disorder, because we are not able to change and control the disorder in given structure.

We are aware that the larger permittivity of HgTe and CdTe in comparison with graphene should substantially reduce the renormalization of the spectrum in HgTe QW structures. On the other hand, the single-electron spectrum of quantum wells at width greater than $10$~nm radically differs from the graphene spectrum. The valence band in these structures consists of two single-spin valleys with a large effective mass, which maxima are significantly shifted from $k = 0$ \cite{Minkov17}, where the bottom of the conduction band is located. The role of such a feature in the renormalization of the spectrum due to many-body  effects remains unclear yet.

\section{Conclusion}

We have presented the results of the detailed, systematic experimental studies of the electron effective mass and its dependence on the electron density in the HgTe quantum wells of different width both for $d<d_c$ where spectrum is normal and for $d>d_c$ where spectrum is inverted. Comparison of these data with  the widely used $kP$ calculations shows: in the structures  with $d<d_c$, the experimental values of the effective mass  are close to the calculated ones over the whole density range; with the increasing QW width, at $d>(7-8)$~nm, the experimental values of effective mass become noticeably less than the calculated ones. This difference increases with the electron density decrease, i.e., with lowering the Fermi energy;  the maximal  difference between the theory and experiment  is achieved at $d = (15-18)$~nm, where the $m_\text{calc}$ to $m_\text{exp}$ ratio reaches the  value of two and begins to decrease with a further $d$ increase. We  assume that observed behavior of the electron effective mass results from the spectrum renormalization by electron-electron interaction.

\acknowledgements

We are grateful to I.V.~Gornyi for useful discussions.
The work has been supported in part by the Russian Foundation for Basic
Research (Grants No. 16-02-00516 and No. 18-02-00050), by  Act 211 Government of the Russian Federation, agreement No.~02.A03.21.0006,  by  the Ministry of Education and Science of the Russian Federation under Project No.~3.9534.2017/8.9, and by the FASO of Russia (theme ``Electron'' No. 01201463326).


%

\end{document}